# Ambipolar Field Effect in Topological Insulator Nanoplates of $(Bi_xSb_{1-x})_2Te_3$

Desheng Kong[1,+], Yulin Chen[2,3,4,+], Judy J. Cha[1], Qianfan Zhang[1], James G. Analytis[2,4], Keji Lai[2,3], Zhongkai Liu[2,3,4], Seung Sae Hong[2], Kristie J. Koski[1], Sung-Kwan Mo[5], Zahid Hussain[5], Ian R. Fisher[2,4], Zhi-Xun Shen[2,3,4], and Yi Cui[1,*]

[1]Department of Materials Science and Engineering, Stanford University, Stanford, California 94305, USA, [2]Department of Applied Physics, Stanford University, Stanford, California 94305, USA, [3]Department of Physics, Stanford University, Stanford, California 94305, USA, [4]Stanford Institute for Materials and Energy Sciences, SLAC National Accelerator Laboratory, 2575 Sand Hill Road, Menlo Park, California 94025, USA, [5]Advanced Light Source, Lawrence Berkeley National Laboratory, Berkeley, California 94720, USA. [+]These authors contributed equally to this work.

Topological insulators represent a new state of quantum matter attractive to both fundamental physics and technological applications such as spintronics and quantum information processing[1-11]. In a topological insulator, the bulk energy gap is traversed by spin-momentum locked surface states forming an odd number of surface bands that possesses unique electronic properties. However, transport measurements have often been dominated by residual bulk carriers from crystal defects[6] or environmental doping[12-14] which mask the topological surface contribution. Here we demonstrate $(Bi_xSb_{1-x})_2Te_3$ as a tunable topological insulator system to manipulate bulk conductivity by varying the Bi/Sb composition ratio. $(Bi_xSb_{1-x})_2Te_3$ ternary compounds are confirmed as topological insulators for the entire composition range by angle resolved photoemission spectroscopy (ARPES) measurements and *ab initio* calculations. Additionally, we observe a clear ambipolar gating effect similar to that observed in graphene[15] using nanoplates of $(Bi_xSb_{1-x})_2Te_3$ in field-effect-transistor (FET) devices, The manipulation of carrier type and concentration in topological insulator nanostructures demonstrated in this study paves the way for implementation of topological insulators in nanoelectronics and spintronics.

Recently, binary sesquichalcogenides $Bi_2Te_3$, $Sb_2Te_3$ and $Bi_2Se_3$ have been identified as three-dimensional (3D) topological insulators with robust surface states consisting of a single Dirac cone in the band spectra[7-9]. In these materials, topological surface states have been experimentally confirmed with surface sensitive probes such as ARPES[8,9,16,17] and scanning tunneling microscopy/spectroscopy (STM/STS)[18,19].. Several recent magnetotransport experiments have also revealed charge carriers originating from surface states[13,20-22]. However, despite the substantial efforts in material doping[9,10,13] and electric gating[23,24], manipulating and suppressing the bulk carriers of these materials, especially in nanostructures, are still challenging owing to impurities formed during synthesis and extrinsic doping from exposure to the ambient environment[12-14].

*To whom correspondence should be addressed: yicui@stanford.edu (Y.C.)

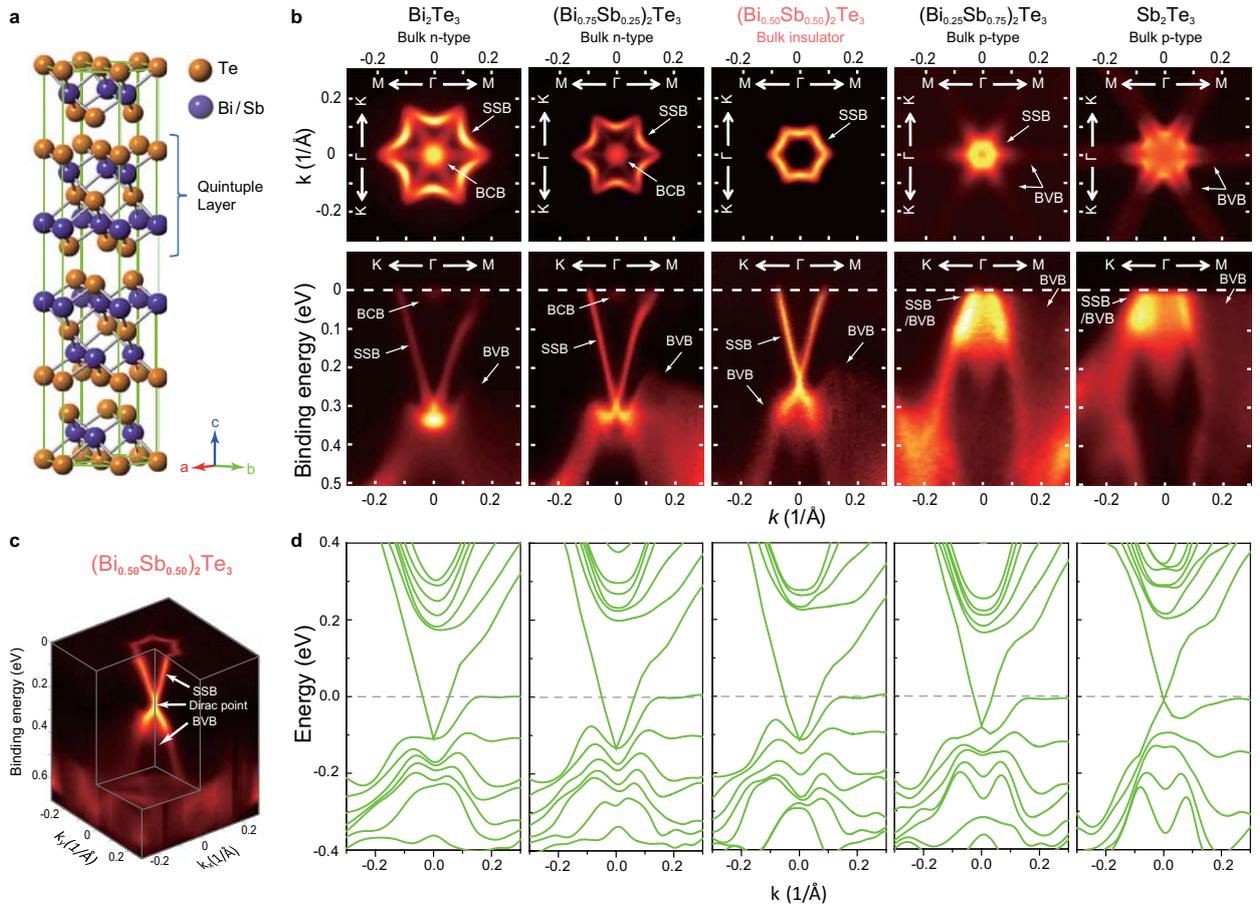

**Figure 1. $(Bi_xSb_{1-x})_2Te_3$ is a tunable topological insulator system with a single Dirac cone of surface states. a,** Tetradymitetype type crystal structure of $(Bi_xSb_{1-x})_2Te_3$ consists of quintuple layers (~ 1 nm in thickness) bonded by Van der Waals interactions. **b,** ARPES Fermi surface (FS) map (top row) and band dispersion along K-Γ-M (bottom row) directions from bulk single crystals with nominal compositions of $Bi_2Te_3$, $(Bi_{0.25}Sb_{0.75})_2Te_3$, $(Bi_{0.50}Sb_{0.50})_2Te_3$, $(Bi_{0.75}Sb_{0.25})_2Te_3$ and $Sb_2Te_3$. By increasing Sb concentration, Fermi energy ($E_F$) exhibits systematic downshift from the bulk conduction band (BCB) to the bulk valence band (BVB) through a bulk insulating state achieved in $(Bi_{0.5}Sb_{0.5})_2Te_3$. The surface state band (SSB) consists of a single Dirac cone around the Γ point, forming a hexagram FS (top row) and V-shape dispersion in the band structure (bottom row). The apex of the V-shaped dispersion is the Dirac point. Note the shape of the Dirac cone (especially the geometry below the Dirac point, which hybridizes with the BVB) also varies with the Bi/Sb composition. For Bi:Sb ratio less than 50:50, as-grown materials become p-type; the $E_F$ resides below the Dirac point thus only the lower part of the Dirac cone is revealed in the ARPES measurement (while the V-shape dispersion inside bulk gap is not seen). The n-type SSB pocket on FS shrinks with increasing Sb concentration and eventually becomes a p-type pocket hybridized with the bulk band (BVB) in the Bi:Sb concentrations of 25:75 and 0:100. **c,** 3D illustration of the band structure of $(Bi_{0.50}Sb_{0.50})_2Te_3$ with vanished bulk states on FS. SSB forms a single Dirac cone with hexagram FS. **d,** Corresponding band structure calculations near Γ point show qualitative agreement with ARPES measurements with gapless SSB consist of linear dispersions spanning the bulk gap are observed in all the compositions. The difference of $E_F$ between calculated and measured band structures reflects the carriers arising from defects and vacancies in the crystals.

Here we propose ternary sesquichalcogenide $(Bi_xSb_{1-x})_2Te_3$ as a tunable 3D topological insulator system allowing us to engineer the bulk properties via the Bi/Sb composition ratio. $(Bi_xSb_{1-x})_2Te_3$ is a non-stoichiometric alloy sharing similar tetradymite structure with the parent compound $Bi_2Te_3$ and $Sb_2Te_3$ (Fig. 1a). To verify the topological nature of $(Bi_xSb_{1-x})_2Te_3$, we performed ARPES measurements on the (0001) plane of $(Bi_xSb_{1-x})_2Te_3$ bulk single crystals of multiple compositions yielding experimental Fermi surface (FS) topology maps and band dispersions (Fig. 1b). Along with the broad electronic spectra originating from the bulk states, the single Dirac cone that forms the topological surface state band (SSB) is revealed around the Γ point for all ternary compositions indicated by the hexagram FSs (top row) and the sharp linear dispersion in the band spectra (bottom row). The parent compounds, as-grown $Bi_2Te_3$ (n-type) and $Sb_2Te_3$ (p-type), are highly metallic with Fermi energy, $E_F$, located deep inside the bulk conduction band (BCB) and bulk valance band (BVB), respectively, due to excessive carriers arising from crystal defects and vacancies[9]. With increasing Sb concentration, $E_F$ systematically shifts downward (Fig 1b, bottom row). In particular, at a Bi/Sb ratio of 1:1, i.e. $(Bi_{0.50}Sb_{0.50})_2Te_3$, the bulk states completely disappear at $E_F$ with a vanished bulk pocket in the FS map and band dispersion. The surface Dirac cone of $(Bi_xSb_{1-x})_2Te_3$ (Fig. 1c) noticeably exhibits clear hexagonal warping, similar to that of pure $Bi_2Te_3$[9,18,25].

The experimentally observed ARPES measurements are qualitatively reproduced by *ab initio* band structure calculations (Fig. 1d) in which the linear SSB dispersion around the Γ point in all compositions confirms their topological non-triviality. This is not surprising as the spin-orbit coupling strength (critical for the formation of topological insualtors[1-3]) and the bulk energy gap in ternary $(Bi_xSb_{1-x})_2Te_3$, with varying Bi/Sb ratios, are comparable to the parent compounds $Bi_2Te_3$ and $Sb_2Te_3$. Consequently, a quantum phase transition to an ordinary insulator does not occur with varying Bi/Sb ratios, complementary to a recent study on the topological trivial/non-trivial phase transition in the $BiTl(S_{1-\delta}Se_\delta)_2$ system[26]. This non-triviality across the entire compositional range in ternary $(Bi_xSb_{1-x})_2Te_3$ compounds demonstrates a rich material assemblage of topological insulators based on an alloy approach, which is an attractive avenue to search for material candidates with improved properties.

The compositional engineering of the bulk properties of topological insulators can also be applied to nanostructures. Single-crystalline $(Bi_xSb_{1-x})_2Te_3$ nanoplates are synthesized by means of a catalyst-free vapor-solid (VS) growth using a mixture of $Bi_2Te_3$ and Sb$_2$Te$_3$ powders as precursors. The growth method has been established in our previous work[27]. Figure 2a shows a typical optical microscopy image of as-grown (Bi$_x$Sb$_{1-x}$)$_2$Te$_3$ nanoplates on an oxidized silicon substrate (300nm SiO$_2$/Si) possessing thicknesses of a few nanometers and lateral dimensions of micrometers. On these substrates, thin layers of nanoplates are semitransparent and can be readily identified with thickness-dependent color and contrast[27] resembling the optical properties of graphene. The single-crystalline nature of these nanoplates is revealed by the clear lattice fringes in high resolution transmission electron microscopy (TEM) images (Fig. 2b) and the sharp selected area electron diffraction spot pattern (Fig. 2c). Energy-dispersive X-ray spectroscopy (EDX) elemental mapping reveals Bi, Sb, and Te distributed across the entire nanoplate without detectable phase separation (Fig. 2d). Nanoplate elemental composition is calibrated by EDX spectra (Fig. 2e). To reflect the initial growth conditions, the nanoplates are labelled with nominal compositions. We fabricated six-terminal hall bar devices on thin nanoplates with thicknesses ranging from 5nm to 10nm for transport measurements (shown schematically in the inset of Fig. 2f). In Fig. 2f, we illustrate the dependence of carrier types and areal carrier densities on composition measured by the Hall effect, which is consistent with the trend in bulk crystals. Binary Bi$_2$Te$_3$ (n-doped) and Sb$_2$Te$_3$ (p-doped) nanoplates contain very high carrier densities of ~10$^{14}$ cm$^{-2}$. By adjusting the composition in ternary (Bi$_x$Sb$_{1-x}$)$_2$Te$_3$ nanoplates, the carrier density systemically drops orders of magnitude with the lowest density achieved in (Bi$_{0.50}$Sb$_{0.50}$)$_2$Te$_3$. In addition to the intrinsic defects formed during synthesis, the carrier concentration in chalcogenide topological insulators is often affected by extrinsic dopants contaminating the sample surfaces from atmospheric exposure[12,13]. For example, we have identified water as an effective n-type dopant that is always present in ambient conditions (Supplementary Information, Fig. S5). A systematic approach to modulate the carrier density is therefore essential for nanostructures to achieve low density since the extrinsic doping depends on environmental conditions.

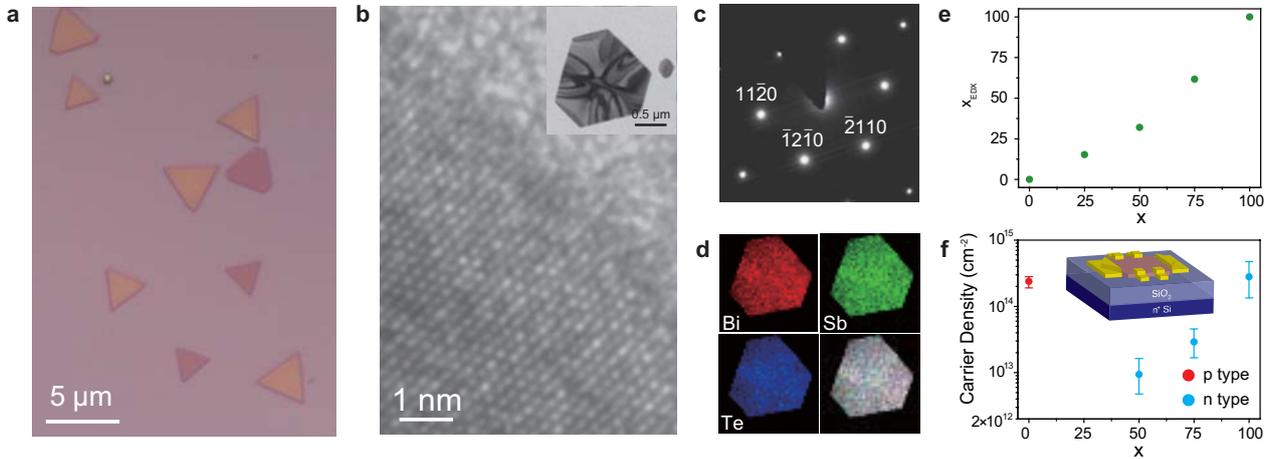

**Figure 2. Characterization of $(Bi_xSb_{1-x})_2Te_3$ nanoplates. a**, Optical microscopy image of vapor-solid grown $(Bi_{0.50}Sb_{0.50})_2Te_3$ nanoplates. **b**, A high-resolution TEM image of the edge of a $(Bi_{0.50}Sb_{0.50})_2Te_3$ nanoplate (shown in the inset) reveals clear crystalline structure with top and bottom surfaces as (0001) atomic planes. **c**, A selected area diffraction pattern with sharp diffraction spots indicates that the nanoplate is a high-quality, single crystal. **d**, Bi, Sb and Te elemental maps obtained from an EDX scan. Overlaying the elemental maps reveals the morphology of the nanoplate, indicating the elements are fairly uniformly distributed without obvious precipitates. **e**, The composition, $x_{EDX}$, in $(Bi_xSb_{1-x})_2Te_3$ nanoplates calibrated by EDX spectra. **f**, The nanoplate area carrier density is chemically modulated by adjusting compositions as determined by the Hall effect. The average carrier concentration from multiple samples is shown as solid circles with error bars corresponding to the maximum deviation. A schematic diagram of the device structure is shown in the inset.

The bulk carriers in low-density (Bi$_x$Sb$_{1-x}$)$_2$Te$_3$ nanoplates can be electrically suppressed with back-gate

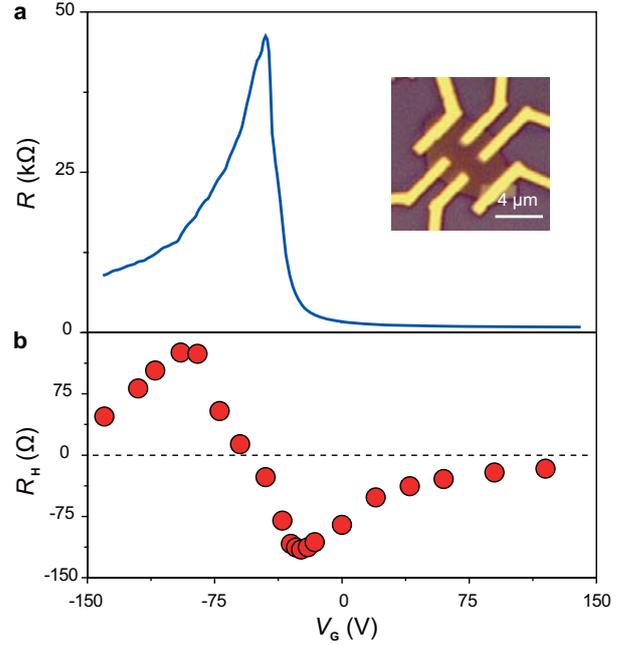

**Figure 3. Ambipolar field effect in a ultrathin nanoplates of (Bi$_x$Sb$_{1-x}$)$_2$Te$_3$. a**, Typical dependence of resistance, $R$, on gate voltage, $V_G$, in a ultrathin (Bi$_{0.50}$Sb$_{0.50}$)$_2$Te$_3$ nanoplate (~5 nm in thickness) exhibiting a sharp peak in the resistance and subsequent decay. Inset: an optical microscopy image of the FET device with a thickness of ~5 nm as determined by AFM. **b**, Hall coefficient, $R_H$, versus, $V_g$, for the same nanoplate. Each $R_H$ (solid circle) is extracted from the Hall trace between ±6T at a certain $V_G$. At the $V_G$ of peak $R$, $R_H$ exhibits a sign reversal.

FET devices. In ultrathin nanoplates (~5 nm), the typical dependence of the resistance, $R$, on the gate voltage, $V_G$, (Fig. 3a) exhibits a very sharp peak that is ~50 times of the resistance at large $V_G$ far from the peak position. The Hall coefficient, $R_H$, reverses its sign when $R$ approaches the maximum value (Fig. 3b). These behaviours resemble the ambipolar field effect observed in graphene[15], which also possesses 2D Dirac fermions. The gate voltage induces an additional charge density and electrostatically dopes the nanoplate altering the nanoplate from an n-type conductor to a p-type conductor through a mixed state in which both electrons and holes are present. For regions with only electrons or holes, $R$ and |$R_H$| decrease with increasing gate-voltage-induced carrier concentration. In the mixed state, $R$ approaches a peak value where the total carrier density is minimized; $R_H$ changes sign when the dominant carrier type (electrons or holes) switches. As expected, there is no zero-conductance region observed presumably due to the presence of surface states inside the bulk bandgap although other contributions cannot be excluded (Supplementary Information, Fig S7 and S8).

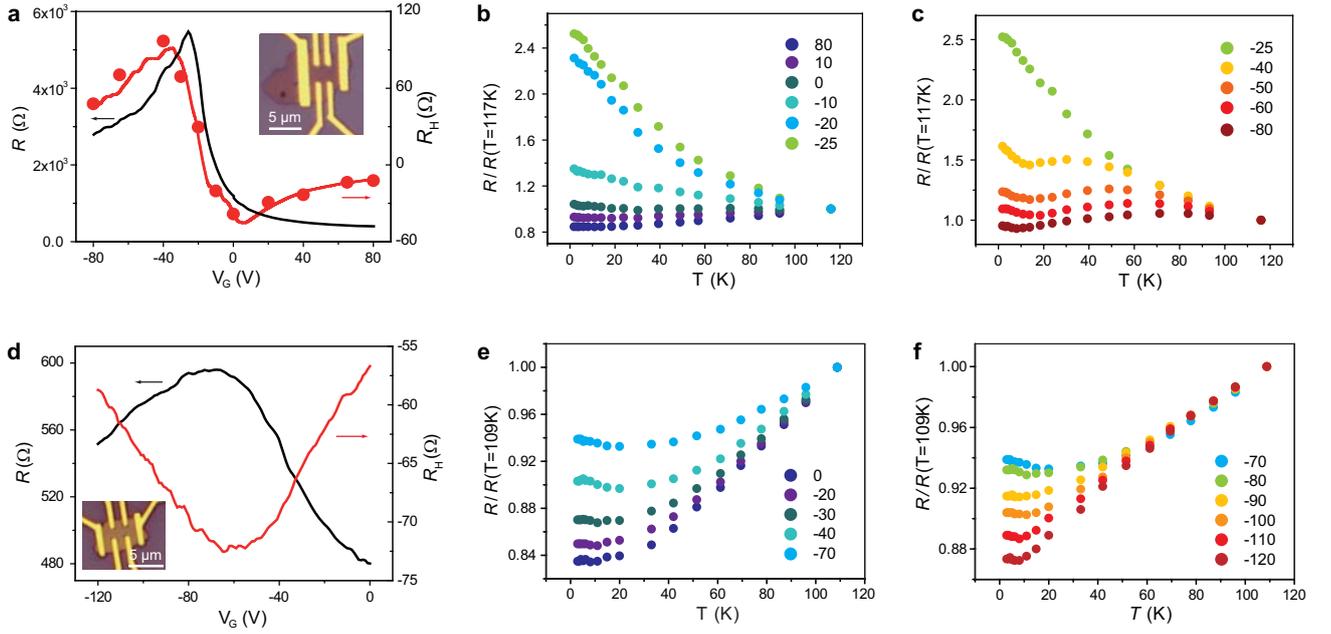

**Figure 4. Temperature-dependent field effect in $(Bi_xSb_{1-x})_2Te_3$ nanoplates. a,** Dependence of resistance, $R$, and Hall coefficient, $R_H$, on gate voltage, $V_G$, from a 5 nm-thick $(Bi_{0.50}Sb_{0.50})_2Te_3$ nanoplate (Inset) showing the ambipolar field effect. For Hall effect measurements, the solid circles are extracted from Hall traces between ±6T at specific $V_G$. The curve is obtained by measuring the Hall resistance versus $V_G$ at magnetic fields of ±4T. **b, c,** Temperature dependence of $R$ at different $V_G$ from electron conductor to mixed state (**b**) and hole conductor to mix state(**c**) respectively. R is normalized to its value at the highest measured temperature. **d,** Dependence of resistance, $R$, and Hall coefficient, $R_H$, on gate voltage, $V_G$, from a 9 nm-thick $(Bi_{0.50}Sb_{0.50})_2Te_3$ nanoplate (Inset). **e, f,** Temperature dependence of R at different $V_G$. R is normalized to the value at the highest measured temperature.

The temperature dependence of $R$ further confirms the suppression of bulk conduction in the mixed state. Systematic dependence studies were performed on another device exhibiting the ambipolar field effect (Fig. 4a). In the purely electron-conductive region, the nanoplate shows typical metallic behaviour with decreasing $R$ as the temperature, $T$, is decreased due to electron-phonon scattering (Fig. 4b). As the nanoplate approaches the mixed state, $R$ begins to increase with continued decreasing $T$ for the entire temperature range (~2 to 120K) primarily due to the freeze-out of bulk carriers, also observed in lightly doped topological insulator $Bi_2Se_3$ crystals[13]. Note that the actual activation energy, $E_a$, cannot be simply extracted by the relation $R \approx R_0 e^{E_a/k_BT}$ ($k_B$ is the Boltzmann constant) owing to multiple-channel conduction in the presence of surface carriers[22]. In addition, the resistance does not diverge at low temperature but gradually saturates to a finite value below 10K consistent with metallic surface conduction in parallel with bulk states. Further sweeping $V_G$ to negative values restores the metallic behaviour of the nanoplate as a hole conductor (Fig. 4c).

Finally, we observe a thickness dependence of the transport measurements in FET devices. The suppression of bulk conduction requires the nanoplate to be much thinner than the depletion length, $D$, the length scale controlled by the gate. An order-of-magnitude estimation found by solving the Poisson equation yields $D \sim 11$ nm (Supplementary Information, Fig S7). For a thicker nanoplate of ~9 nm in thickness, however, the dependence of $R$ on $V_G$ (Fig. 4d) shows a much weaker tenability than the ultrathin nanoplates, and the entire sample remains n-type as $R_H$ does not reverse the sign. In addition, $R$ decreases with $T$ for all $V_G$ exhibiting metallic behaviour until ~20K below which weak carrier freeze-out is indicated by a slight rise in $R$ (Fig. 4e-f). Apparently, the nanoplate is too thick to be effectively depleted by gating, and large metallic bulk conduction is always present in the device and contributes to the charge transport. It is therefore important to fabricate FET devices from few-layer nanoplates for the effective manipulation of bulk conductivity. We also note that in the thinnest limit, the top and bottom surface states may hybridize by quantum tunnelling and open an energy gap, resulting in either a conventional insulator or two dimensional quantum spin hall system[19,28-30]. The measured thin nanoplates (≥5nm) are beyond the thickness threshold for such a transition.

## METHODS

Synthesis. Single crystals of $Bi_xSb_{2-x}Te_3$ were obtained by slow cooling a binary melt of varying Bi/Sb/Te ratios. This mixture was sealed in quartz under a partial pressure of argon. The mixture was heated to a temperature of 800 °C over 14 hrs, held for an additional 6 hrs, then cooled to 500 °C for 100 hrs, and finally the furnace was naturally cooled to room temperature.

Ultrathin $Bi_xSb_{2-x}Te_3$ nanoplates were grown inside a 12 inch horizontal tube furnace (Lindberg/Blue M) with a 1-inch diameter quartz tube. A uniform mixture of $Bi_2Te_3$ and $Sb_2Te_3$ powders (Alpha Aesar, 99.999%) with specific molar ratio is placed at the hot center region as precursors for evaporation. Degenerately doped silicon substrates with 300nm thermally grown oxide film were placed downstream at certain locations using the temperature gradient along the tube to control the growth temperature. The tube was initially pumped down to a base pressure less than 100 mtorr and flushed with ultrapure argon repeatedly to reduce residual oxygen. During growth, argon flow provided impetus to transport the vapor to the subtrates. The typical growth conditions of $(Bi_xSb_{1-x})_2Te_3$ nanoplates are: ~0.5 g power mixture, 10 torr pressure, 15 s.c.c.m. carrier gas flow, 490 °C precursor temperature, 10 min duration time. $(Bi_xSb_{1-x})_2Te_3$ tended to grow at locations of ~12cm away from the hot centre region, corresponding to a temperature of ~300 °C.

Angular Resolved Photoemission Spectroscopy Measurements. ARPES measurements were performed at beamline 10.0.1 of the Advanced Light Source (ALS) at Lawrence Berkeley National Laboratory. The measurement pressure was kept below $3\times10^{11}$ torr for all time, and data was acquired by Scienta R4000 analyzers at a 10K sample temperature. The total convolved energy and angle resolutions were 15meV and 0.2°, i.e. 0.012(1/Å) for photoelectrons excited by 48eV photons. The fresh surface for ARPES measurement was obtained by cleaving the sample *in situ* along its natural cleavage plane.

Theoretical calculations. The first-principle electronic band calculations were performed in 6-quintuple-layer slab geometry using the Vienna *Ab-initio* Simulation Package (VASP). The Perdew-Burke-Ernzerhof type generalized gradient approximation was used to describe the exchange–correlation potential. SOC was included using scalar-relativistic eigenfunctions as a basis after the initial calculation is converged to self-consistency. A k-grid of 10×10×1 points was used in the calculations and the energy cutoff set to 300 eV. A 2 × 2 unit cell of Bi and Sb atoms randomly substituted is used to simulate the structure. Lattice parameters of $(Bi_xSb_{1-x})_2Te_3$ are interpolated from experimental lattice parameters of $Bi_2Te_3$ and $Sb_2Te_3$ according to the composition.

Nanostructure Characterizations. Characterization was done using optical microscopy (Olympus BX51M, imaged with 100X objective under normal white illumination), TEM (FEI Tecnai G2 F20 X-Twin microscope, acceleration voltage 200kV) equipped with an energy dispersive X-ray spectrometer, and AFM (Park Systems XE-70). For TEM and EDX characterizations, $(Bi_xSb_{1-x})_2Te_3$ nanoplates are directly grown on 50nm $Si_3N_4$ membranes supported by silicon windows. The actual composition of $(Bi_xSb_{1-x})_2Te_3$ nanoplates are calibrated by EDX spectra with $Bi_2Te_3$ and $Sb_2Te_3$ spectra as references.

Device Fabrication. Back-gate FET devices were directly fabricated on as-grown substrates with 300nm $SiO_2$ films on silicon. The substrates were first decorated with metal markers based on standard e-beam lithography followed by thermal evaporation of Cr/Au (5nm/60nm). After a suitable nanoplate was selected, a second patterning step defined multiple Cr/Au (5nm/100nm) electrodes by means of the markers.

Transport Measurements. Low-frequency (~ 200Hz to 1000Hz) magnetotransport experiments were carried out in an Oxford cryostat with digital lock-in amplifiers (Stanford Research Systems SR830). All transport measurements are measured at the base temperature of 2.0K unless specified otherwise. A Keithley 2400 sourcemeter was used to apply gate voltage. The resistance is measured with standard four-terminal configuration to eliminate contact resistance.

*Acknowledgement.* Y.C. acknowledges the supported from the Keck Foundation, and King Abdullah University of Science and Technology (KAUST) Investigator Award (No. KUS-l1-001-12). Y.

L. C., Z.K.L., Z.X.S., J.G.A. and I.R.F acknowledge the support from Department of Energy, Office of Basic Energy Science, under contract DE-AC02-76SF00515. K.L. acknowledges the KAUST Postdoctoral Fellowship support No. KUS-F1-033-02.

*Author contributions.* D.K., Y.L.C. and Y.C. conceived the experiments. Y.L.C and Z.K.L. carried out ARPES measurements. J.G.A. synthesized and characterize bulk single crystals. Q.F.Z. performed electronic structure calculations. D.K. and J.J.C. carried out synthesis, structural characterization and device fabrication for nanoplates. D.K., K.L., J.J.C., S.S.H. and K.J.K. carried out transport measurements and analyses. All authors contributed to the scientific planning and discussions.

*Additional information.* The authors declare no competing financial interests. Supplementary information accompanies this paper at arxiv.org. Correspondence and requests for materials should be addressed to Y.C, email: yicui@stanford.edu.